# Design and Implementation of an IoT-based Respiratory Motion Sensor

Bardia Baraeinejad, Maryam Forouzesh, Saba Babaei, Yasin Naghshbandi, Yasaman Torabi, and Shabnam Fazliani

*Abstract*— In the last few decades, several wearable devices have been designed to monitor respiration rate in an effort to capture pulmonary signals with higher accuracy and reduce patients' discomfort during use. In this article, we present the design and implementation of a device for real-time monitoring of respiratory system movements. When breathing, the circumference of the abdomen and thorax changes; therefore, we used a Force Sensing Resistor (FSR) attached to the Printed Circuit Board (PCB) to measure this variation as the patient inhales and exhales. The mechanical strain this causes changes the FSR electrical resistance accordingly. Also, for streaming this variable resistance on an Internet of Things (IoT) platform, Bluetooth Low Energy (BLE) 5 is utilized due to the adequate throughput, high accessibility, and possibility of power consumption reduction. Furthermore, this device presents features such as low power consumption (400 µW), high precision, and ease of use.

## I. INTRODUCTION

RESPIRATORY rate frequency provides additional levels of information about the physical and mental condition of a person. Therefore, accurate monitoring of the respiratory system allows for a more precise diagnosis for people with certain medical conditions [1] [2]. Abnormalities in the respiratory cycle are closely linked to various respiratory diseases, emphasizing the importance of understanding its dynamics [3]. The respiratory cycle consists of two main phases: inspiration and expiration. During inspiration, the lungs expand to draw in environmental air, while expiration involves the relaxation of the lungs to release a gas mixture, primarily composed of carbon dioxide [4] [5]. In regard to the physical problems prediction, we can mention the application of respiration monitoring sensors in recognizing signs of respiratory depression in patients recovering from surgery and anesthesia, detecting sleep-related breathing disorders, and, most recently, for managing patients with COVID-19 to constantly tracking pulmonary function [1] [6] [7]. More to the point, respiration rate can also be associated with the emotional state of a person, for instance, sadness, depression, happiness, anxiety, etc. [8]. Considering the vital role of proper functioning of the respiratory system, keeping track of a patient's breathing rate noninvasively has always been an issue of interest. Also, the design of these sensors must minimize the impact on the patient's daily life and provide comfort during use [9]. A variety of research has been done on developing respiratory monitoring sensors. To name but a few, bed-type sleep monitoring systems have been proposed, which can be used only during sleep. Furthermore, some non-contact wearable sensors have recently been developed, which may also cause discomfort for patients due to the additional belts attached [10]. In this work, we designed a handy, easy-to-use Printed Circuit Board (PCB) containing a Force Sensitive Resistor (FSR) to measure the pressure and an accelerometer that ultimately senses the respiration signal of a patient. We can measure force by measuring the changes in FSR resistance. It is also possible to detect the motion of the body with the help of a 3-axis accelerometer [11]. Further benefits of incorporating FSR and accelerometer include accuracy, cost-effectiveness, and thinness [12]. Altogether, we have developed a wearable device for capturing respiratory signals with a convenient and unobstructed design for patient use. In addition, we have offered features, including high precision and power consumption of 4.9 mW.

## II. HARDWARE DESIGN

### A. Hardware Overview

The goal of this research is to design a device that can measure the respiration rate through the measurement of chest movements. The components of this device had been selected according to their low power consumption, functionality, and high relative accuracy.

As depicted in the hardware diagram (Fig. 2), an ARM Cortex M4F processor and various peripherals, such as Bluetooth Low Energy (BLE) 5, are deployed [13]. Additionally, there is a USB Type C jack that is connected to the battery charger (MCP73833T) to charge a 450 mAh Lithium Polymer (LiPo) battery [14]. LiPo batteries can meet the needs of medical device applications due to their remarkable features like good cycle life, rechargeability, lightweight, and high energy density [15]. It has an accelerometer (LIS2DH12) for detecting the user motion [11]. A high-efficiency miniature DC-DC switching buck converter must be used to supply 1.8 V from the LiPo battery to the FSR, Micro-Controller Unit (MCU), and related circuits. TPS62840 is an excellent choice for this goal. It also has enough load capability [16].

### B. MCU

We used nRF52832 as the MCU. It has a low-power 32-bit ARM Cortex M4F processor. Also, BLE 5, which has enough throughput and low power consumption, exists in nRf52832 as a peripheral. The processor has other peripherals as well as General Purpose Input/Outputs (GPIOs) and an Analog to Digital Converter (ADC). Both battery percentage and FSR are measured using ADC.

### C. FSR

An FSR, whose electrical resistance decreases the more force is applied to its surface, is used in this work. FSR has some advantages over strain gauges or load cells, like cost efficiency, low profile, and excellent force mapping application (changes occur significantly when the pressure is applied to the FSR surface), and also ensures patient comfort [17] [18] [19]. For further comparison of force-sensing devices and why FSR was explicitly chosen, refer to the Comparison section and Table 1. As shown in Fig. 1, to measure the force applied, a voltage divider is implemented by placing a fixed resistor (499 kΩ) in a series with the FSR, which acts as a variable resistance. By increasing the pressure on the FSR, the resistance of the FSR decreases. The output voltage would be obtained by the equation 1.

$$V_{out} = V_{dd} \times \frac{R}{(R + R_{FSR})} \quad (1)$$

With the addition of the ADC, we are able to sample the electrical resistance, thereby measuring the force.

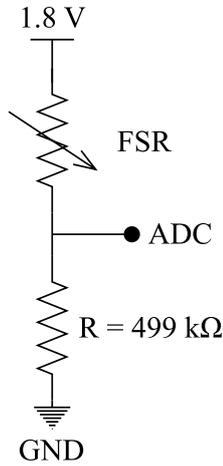

**Fig.1** Reading/measuring FSR resistance

### D. Power Management

To supply the device, we use a rechargeable 450 mAh Lithium Polymer (LiPo) battery. A LiPo battery has been chosen for our device due to its high energy density, rechargeability, and decent cycle life. A further advantage of these batteries is their light weight, flexibility, and leak resistance. Also, we incorporate a buck converter (TPS62840) to regulate the battery voltage (3.3 V ~ 4.2 V) to 1.8 V in order to supply the elements that require this voltage. Using a switching converter enhances the battery life and efficiency [15]. A battery charger (MCP73833T) charges the battery through a USB type-C connector [14]. Moreover, an ADC measures the battery charge every 2 s. Then, through a resistor divider, the battery voltage is scaled to the microcontroller voltage for the purpose of sending a percentage level and ultimately charging status via BLE.

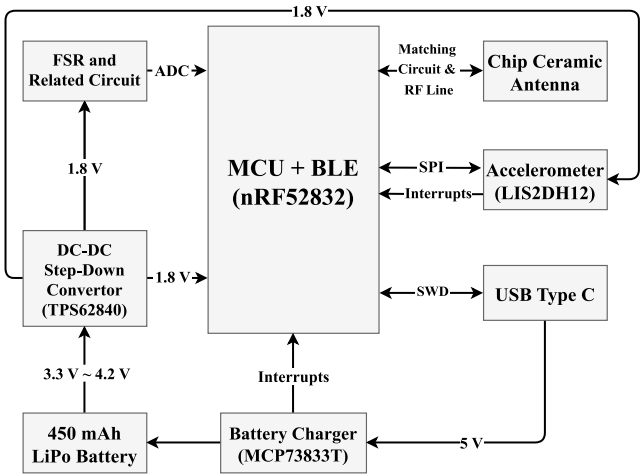

**Fig. 2.** Block diagram of the Device. The common supply voltage for most of the components is 1.8 V. However, the Step-Down Converter is directly powered by a 450 mAh LiPo battery.

### E. Accelerometer

LIS2DH12 is a low-power, high-performance 3-axis accelerometer that detects motion, tracks physical activity, and identifies unexpected movements [11].

### F. Firmware

As shown in Fig. 3, the data is sent through the device to a smartphone using BLE 5. It also indicates the battery percentage. The first channel of the ADC measures the battery percentage, and the second channel measures the FSR pressure. The Accelerometer streams the data at 50 samples per second.

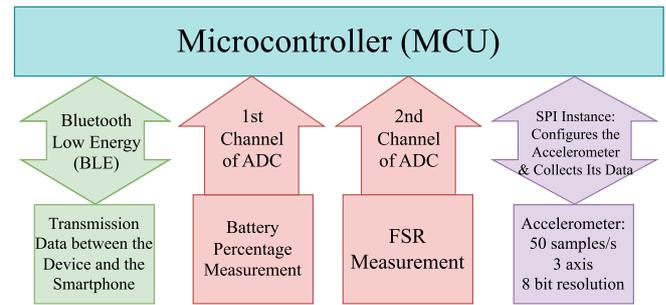

**Fig. 3.** Firmware flowchart depicting the overall functionality. SPI configures the accelerometer. The first channel of ADC is used for the battery percentage measurement, and the second is for the FSR measurement. Also, the firmware is responsible for data transmission through BLE between the device and the smartphone.

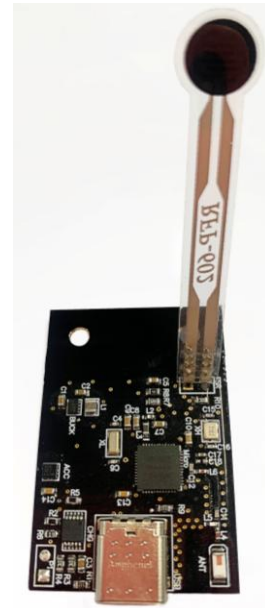

**Fig 4.** PCB of the respiration device

### G. Body Design

The enclosure of this device is manufactured by Stereolithography (SLA) resin printing according to its high accuracy, isotropic, high thermal durability, and high finish surface [20]. The material used for printing is ABS-like (Acrylonitrile Butadiene Styrene) resin. Although it might be a little more expensive than standard resin, it has more impact resistance, tensile strength, and durability [21] [22]. The implementation of the respiration monitoring device succeeds in delivering a sturdy, user-friendly, and efficient practical device.

## III. Comparison with Existing Systems

Several devices have been developed to measure the respiration rate in a variety of designs, including skin electrodes, respiration belts, and sleep-related respiratory sensors [9] [10]. However, the advantages of good comfort, a high Signal-to-Noise Ratio (SNR), the possibility of real-time monitoring and long-term detection of biological signals make sensors placed directly on the skin a suitable choice for respiratory sensors, among other types [23].

There are also several technologies for measuring pressure, including load cells, strain gauges, and FSRs. Each of these approaches has its pros and cons, and one must choose the one that produces the best results in terms of accuracy, efficiency, and cost [24]. In our work, we used FSR for pressure sensing. Comparing devices using FSRs with ones using strain gauges, we find that when the same force is applied to the sensor, there is a broader swing in the FSR's output than in strain gauges [25]. As an additional advantage, FSR technology has overload cells and strain gauges with its minimally invasive form factor, thin shape, and capability to cover a wider sensing region if needed [25]. As stated above, as a result of the dramatic changes in FSR resistance of FSRs with pressure, any force applied due to respiration can be easily observed [17]. Based on all these considerations, we decided to implement FSR into our wearable sensing device.

TABLE I

| | Advantages | Limitations |
| --- | --- | --- |
| FSR | Small size, low cost, good shock resistance and linear transfer function for small forces, high durability, no battery usage, capable of measuring a wide range of forces because they can be manufactured in different shapes and force ranges [26] [27] [28] | Non-negligible hysteresis for high force values [28] |
| Strain Gauge | Small size, multi-axis measurement, resistant to temperature changes, accuracy and linearity [29] | Sensitive to electromagnetic noise and changes in temperature [29] |
| Load Cell | Accuracy and reliability [30] | Bulky in size, rigid in construction, expensive electronic, and higher power consumption than FSRs [26] |
| Capacitive Force Sensor | Low power consumption Large bandwidth High sensitivity High resolution [31] [24] [32] [33] | Complex circuit Sensitive to noise Sensitive to temperature [31] [24] [32] [33] Energy Consumption |
| Optical Force Sensor | High sensitivity to changes Flexible High sensitivity [31] [24] [29] | High price Bulky in size Light sensitivity [31] [24] [29] |

SENSORS COMPARISON
A comparison between various sensors and their features.

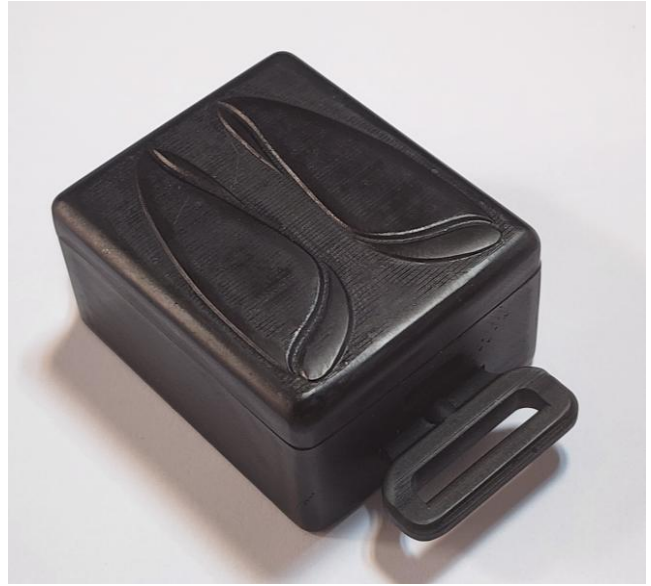

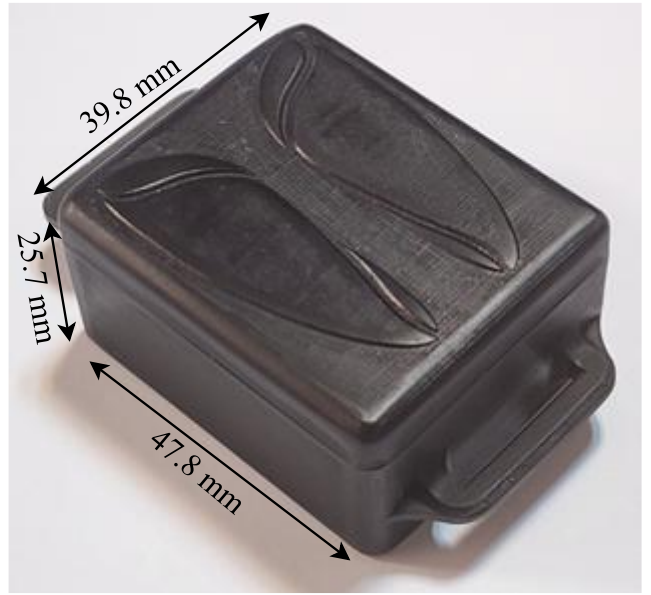

**Fig 5.** Final Body Design

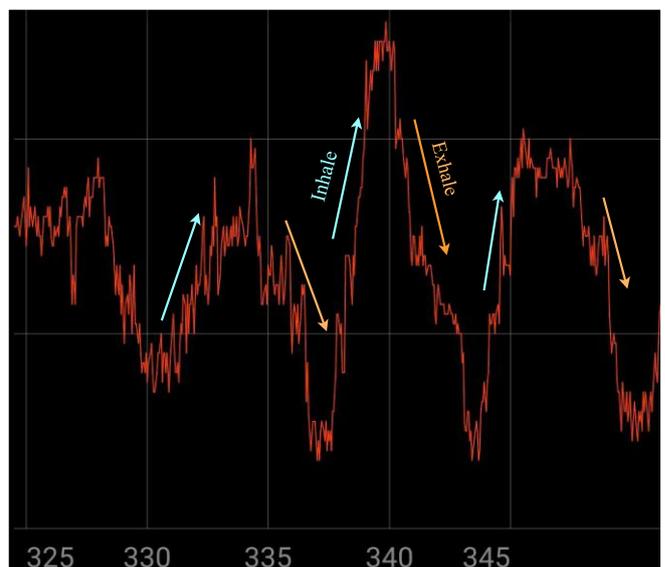

**Fig 6.** Illustrating the Signal During Device Usage

## IV. Conclusion

The designed respiratory monitoring system introduces a number of novel design elements and satisfies the desired functional requirements. The selection of proper elements in the structure of the sensor, as well as innovative design approaches, ensure a high resolution of 12 bits for capturing respiratory signals and long battery life. Also, the size of the device makes it portable and convenient to use for patients. Moreover, it provides a system for sending notifications and communicating between the device and its user.